\documentclass[fleqn,usenatbib]{mnras}

\usepackage{newtxtext,newtxmath}
\usepackage{orcidlink}

\usepackage[T1]{fontenc}

\DeclareRobustCommand{\VAN}[3]{#2}
\let\VANthebibliography\thebibliography
\def\thebibliography{\DeclareRobustCommand{\VAN}[3]{##3}\VANthebibliography}

\usepackage{graphicx}	% Including figure files
\usepackage{amsmath}	% Advanced maths commands

\title[Formation of Be star decretion discs]{Formation of Be star decretion discs through boundary layer effects}

\author[R. G. Martin et al.]{
Rebecca G. Martin,$^{1,2}$\orcidlink{0000-0003-2401-7168}\thanks{E-mail: rebecca.martin@unlv.edu}
Stephen H. Lubow,$^{3}$\orcidlink{0000-0002-4636-7348}
David Vallet,$^{1,2}$\orcidlink{0000-0002-0543-6730}
Madeline Overton,$^{1,2}$\orcidlink{0009-0000-7649-0593}\newauthor
\, Stephen Lepp$^{1,2}$\orcidlink{0000-0003-2270-1310} 
 and Zhaohuan Zhu$^{1,2}$\orcidlink{0000-0003-3616-6822}
\\
$^{1}$Nevada Center for Astrophysics, University of Nevada, Las Vegas,
4505 South Maryland Parkway, Las Vegas, NV 89154, USA\\
$^{2}$Department of Physics and Astronomy, University of Nevada, Las Vegas,
4505 South Maryland Parkway, Las Vegas, NV 89154, USA\\
$^{3}$Space Telescope Science Institute, 3700 San Martin Drive, Baltimore, MD 21218, USA
}

\date{Accepted XXX. Received YYY; in original form ZZZ}

\pubyear{\the\year{}}

\begin{document}
\label{firstpage}
\pagerange{\pageref{firstpage}--\pageref{lastpage}}
\maketitle

\begin{abstract}
Be stars are rapidly rotating, with angular frequency around $0.7-0.8$ of their  Keplerian break up frequency, as a result of significant accretion during the earlier stellar evolution of a companion star. Material from the equator of the Be star is ejected and forms a decretion disc, although the mechanism for the disc formation has remained elusive. We find one-dimensional steady state decretion disc solutions that smoothly transition from a rapidly rotating star that is in hydrostatic balance.  Boundary layer effects in a geometrically thick disc which connects to a rotationally flattened star enable the formation of  a decretion disc at stellar spin rates  below the break up rate. For a disc with an aspect ratio $H/R\approx 0.1$ at the inner edge, the torque from the disc on the star slows the stellar spin to the observed range and mass ejection continues at a rate consistent with observed decretion rates. The critical rotation rate, to which the star slows down to, decreases as the disc aspect ratio increases. More generally, steady state accretion and decretion disc solutions can be found for all stellar spin rates. The outcome for a particular system depends upon the balance between  the decretion rate and any external infall accretion rate.
\end{abstract}


%%%%%%%%%%%%%%%%%%%%%%%%%%%%%%%%%%%%%%%%%%%%%%%%%%

%%%%%%%%%%%%%%%%% BODY OF PAPER %%%%%%%%%%%%%%%%%%

\section{Introduction}

Be stars are B type stars that are rapidly rotating \citep{Slettebak1982,Porter1996} and have shown H$\alpha$ emission \citep{Collins1987,Porter2003}.  Be stars have average angular spin frequencies of around $\left<\Omega_*\right>=0.7-0.8\,\Omega_{\rm *  K}$ \citep[e.g.][]{Rivinius2013}, where $\Omega_{\rm * K}$ is the Keplerian break up frequency. For an isolated star, the break up rotation frequency is $\Omega_{\rm * K}=(G M_*/R_*^3)^{1/2}$, where $M_*$ is the mass of the star and $R_*$ is its radius.  The initial spin of the Be star could be primordial, a result of binary mass transfer or a result of spin-up during the main-sequence B star phase \citep{McSwain2005}. However, Be stars are typically in a binary system  with a neutron star companion \citep[e.g.][]{Raguzova2005,Reig2011} and are thought to have been spun up through the accretion of material during the earlier evolution of the companion star \citep[e.g.][]{Pols1991,deMink2013,Dodd2024}. While stellar winds may cause spin-down of the star \citep{Lau2011,Nathaniel2025},  Be stars are observed at all ages through the main-sequence \citep{Mermilliod1982,Slettebak1985}.

Be stars have a viscous Keplerian disc \citep{Lee1991,Hanuschik1996,Quirrenbach1997,Hummel1998,Okazaki2002,Jones2008, Wheelwright2012,Rivinius2013,Okazaki2016,Franchini2019bestars}. The formation mechanism of the decretion disc \citep{Pringle1991}  has long remained elusive.  Some Be star systems are observed to have stable discs that implies a constant feeding mechanism  while other systems show disc build up and decay phases \citep{Rivinius2013}. Since the supernova explosion that formed the neutron star may be associated with a kick, Be/X-ray binaries are often in eccentric orbits the the spin of the star misaligned to the binary orbit \citep{Brandt1995,Martinetal2009b,Salvesen2020}. A large misalignment between the disc and the binary orbit can also lead to a range of dynamical effects that may cause increased accretion back on to the Be star \citep[e.g.][]{Martinetal2014,Franchini2021,Suffak2022}.

Most theoretical models of the evolution of Be star decretion discs, assume that material is injected into the Be star disc from the equator of the Be star. However, when injection occurs at a small radius with a zero-torque inner boundary, most of the material quickly falls back onto the Be star and only a small amount goes into the decretion disc \citep[e.g.][]{Okazaki2002,Martin2024}. This leads to implausibly large injection accretion rates being required to form the disc \citep{Nixon2020,Nixon2021}.

These issues can be overcome if the star exerts a torque on the disc. \cite{Nixon2020} explored the magnetic truncation of the inner disc at the Alfven radius \citep{Pringle1972}. Provided that the Alfven radius is larger than the corotation radius  of the star and the disc \citep[but not so large as to destroy the disc,][]{udDoula2018}, then angular momentum can be transferred from the star to the inner disc and prevent the reaccretion of material.   In this model, there is gap between the star and the inner radius of the disc.

In this Letter we suggest that the required torque on the disc may be provided by the star in a boundary layer that connects a geometrically thick disc to a rotationally flattened star. 
Boundary layer effects in hydrodynamical accretion discs have long been investigated \citep[e.g.][]{Pringle1977,Pringle1979,Regev1983,Popham1991,Paczynski1991, Colpi1991,Popham1993,Bisnovatyi1993,Regev1995}.  Recently, \cite{Dong2021} used 2D hydrodynamical simulations and found that there is a maximum rotation rate for a planet with a circumplanetary accretion disc of about $0.7-0.8\,\Omega_{\rm * K}$ for disc aspect ratio $H/R\approx 0.1$. Above this value, the disc becomes a decretion disc and angular momentum from the planet is lost to the disc.
With a one-dimensional steady state disc model, we show that similar boundary layer effects may explain the formation of Be star decretion discs.  In Section~\ref{model} we find steady state accretion and decretion disc solutions. In Section~\ref{discstar} we use analytic approximations to the disc solutions in the boundary layer and match the solutions to a model of a rotating star. We find the timescale for the stellar spin to change as a result of the torque from the disc. The spin rate to which the star evolves to over its lifetime is in good agreement with observed Be star spins.  We discuss implications of the model  and draw our conclusions in Section~\ref{concs}.

\section{Steady state disc model}
\label{model}

In this Section we describe the steady state  disc model and find numerical solutions and analytical approximations for the disc structure. 

\subsection{Disc equations}

We follow the one-dimensional model for accretion discs developed in \cite{Pringle1981} but include the boundary layer effects  described in \cite{Popham1991} and \cite{Paczynski1991}. The disc extends from the inner radius $R_*$, that coincides with the radius of the star up to the outer radius $R_{\rm out}$. We leave a discussion of the definition of the stellar radius to the next Section. The constant steady state accretion rate through the disc is
\begin{equation}
    \dot M = -2 \pi R v_{\rm R}\Sigma
    \label{mdot}
\end{equation}
\citep{Pringle1981}, where $\Sigma$ is the disc surface density and $v_{\rm R}$ is the radial velocity. Note that $\dot M$ is defined to be positive for an inflowing accretion disc with $v_{\rm R}<0$ and negative for an outflowing decretion disc with $v_{\rm R}>0$.
The fluid in the disc has angular frequency $\Omega$.

The viscosity is parameterised with
\begin{equation}
    \nu=\alpha \frac{c_{\rm s}^2}{\Omega_{\rm K}}
\end{equation}
\citep{Pringle1981}, where $\alpha$ is the \cite{SS1973} viscosity parameter and the sound speed
is
\begin{equation}
    c_{\rm s}=\left(\frac{H}{R}\right) R \Omega_{\rm K}.
\end{equation}
The Keplerian angular frequency is given by
\begin{equation}
    \Omega_{\rm K}=\left(\frac{G M_*}{R^3}\right)^{1/2}.
    \label{omkep}
\end{equation}
The disc aspect ratio, $H/R$, is taken to be  constant with radius.

 The radial equation of motion is 
\begin{equation}
    \frac{dv_{\rm R}}{dR}-\frac{v_{\rm R}\left[(\Omega_{\rm K}^2-\Omega^2)R^2-c_{\rm s}^2\right]}{R(c_{\rm s}^2-v_{\rm R}^2)}=0
    \label{eqvr}
\end{equation}
\citep[see equation 7 in ][]{Popham1991}.
Note that all solutions we consider are subsonic, $v_{\rm R} \ll c_{\rm s}$. Conservation of angular momentum is
\begin{equation}
    \frac{d\Omega}{dR}-\frac{\Omega v_{\rm R}}{\nu} \left(1-\frac{j R_{*}^2\Omega_{\rm * K}}{R^2\Omega }\right)=0
    \label{eqom}
\end{equation}
\citep[see equation 10 in][]{Popham1991},
where the constant $j$ is given by
\begin{equation}
    j=\frac{\dot J}{\dot M \Omega_{\rm * K} R_*^2},
%    \label{eqj}
\end{equation}
and $\dot J$ is the rate of flow of angular momentum. 
We discuss the value of $j$ further in Section~\ref{j}.

\subsection{Keplerian disc solutions}
\label{Kep}

For comparison, we first find disc solutions in the Keplerian limit ($\Omega=\Omega_{\rm K}$). 
Solving equation~(\ref{eqom}) in the Keplerian limit gives
\begin{equation}
       v_{\rm R,K}=-\frac{3\nu }{2 R }\left[1 - j \left(\frac{R_*}{R}\right)^{1/2}\right]^{-1} .
    \label{vrkep}
\end{equation}
Substituting this into equation~(\ref{mdot}) gives
\begin{equation}
\nu \Sigma = \frac{\dot M }{3\pi}\left[1-j\left(\frac{R_*}{R} \right)^{1/2}\right].
\label{Kepacc}
\end{equation}
With $j=1$, this is the classic steady state surface density profile for an {\it accretion} disc \citep{Pringle1981}. 
Similarly, we can find the classic decretion disc steady state disc if we set $j=j_{\rm dec}$, where
\begin{equation}
    j_{\rm dec}  = \left(\frac{R_{\rm t}}{R_*}\right)^{1/2}.
%    \label{eqj}
\end{equation}
The steady state {\it decretion} disc solution is
\begin{equation}
\nu \Sigma = \frac{(-\dot M) }{3\pi}\left[\left(\frac{R_{\rm t}}{R} \right)^{1/2}-1\right]
\label{Kepdsc}
\end{equation}
\citep[e.g.][]{Carciofi2008}. The radius $R_{\rm t}$ is a radius where the surface density is zero at the outer edge  of the disc. In a time-dependent disc, this radius may increase in time as the disc spreads outwards. In order for a completely steady solution, mass must be removed at this radius. In a binary star system, the decretion disc can reach a steady state value for $j$ since the binary companion can tidally truncate the disc \citep[e.g.][]{Papaloizou1977,Artymowicz1994} even with the mass ratio of a Be/X-ray binary  \citep[e.g.][]{Okazaki2001,Panoglou2018,Martin2024,Rast2024}. However, we do note that the strength of the tidal torque depends on the alignment of the disc to the binary orbital plane  \citep{Lubowetal2015,Miranda2015,Cyr2017,Overton2024}

\subsection{Boundary conditions}

At the inner radius, the disc rotates at the same frequency as the star and so we choose the inner boundary condition 
\begin{equation}
\Omega(R_{*})=\Omega_*.
\end{equation}
The application of this boundary condtion applies a torque on the disc.
In the limit of $R\gg R_{\rm in}$, the disc is Keplerian so that $\Omega=\Omega_{\rm K}$.  We apply an outer boundary condition that the radial velocity is Keplerian so that
\begin{equation}
    v_{\rm R}(R_{\rm out})=v_{\rm R,K}.
\end{equation}
Note that we also tried using an outer boundary condition of $\Omega=\Omega_{\rm K}$. The solutions are unchanged but less numerically stable.

 \begin{figure*}
     \centering
     \includegraphics[width=0.68\columnwidth]{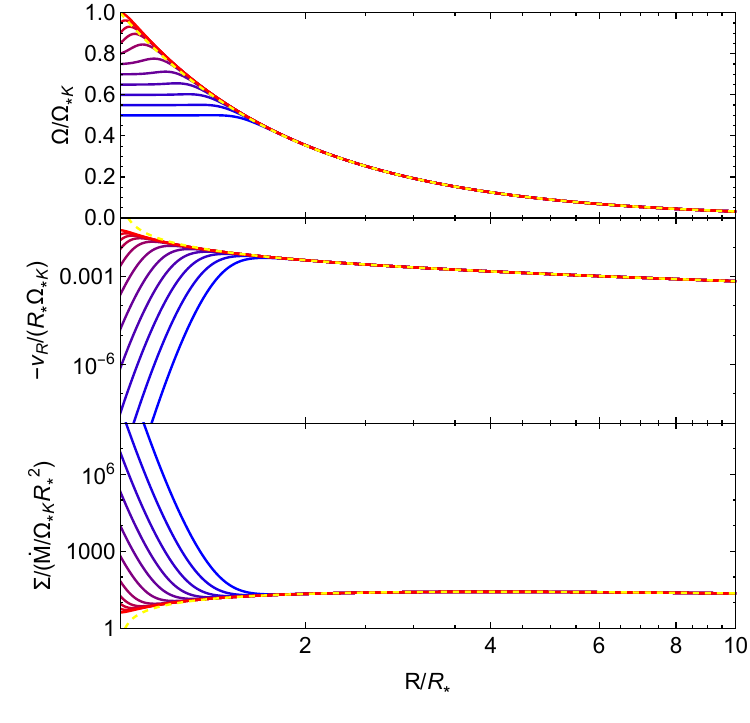}
     \includegraphics[width=0.68\columnwidth]{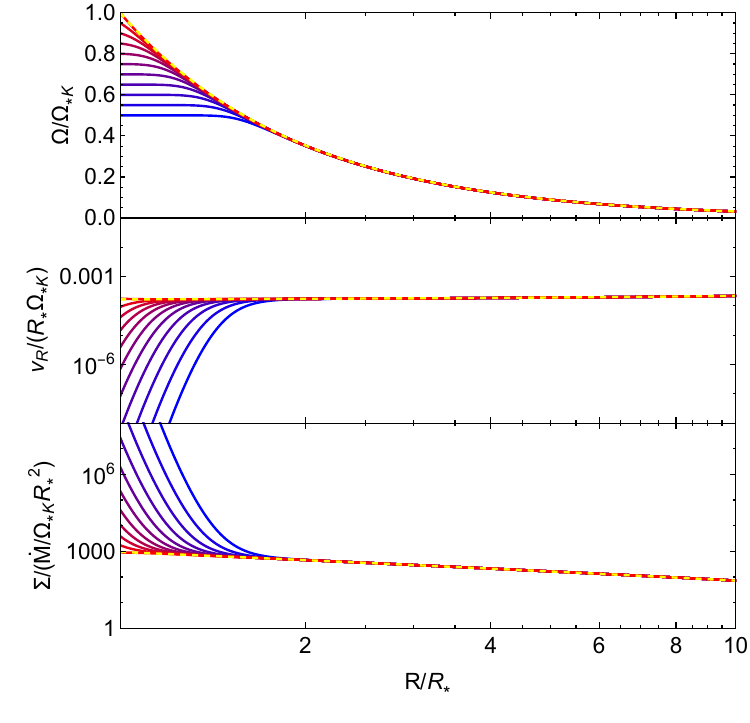}
      \includegraphics[width=0.68\columnwidth]{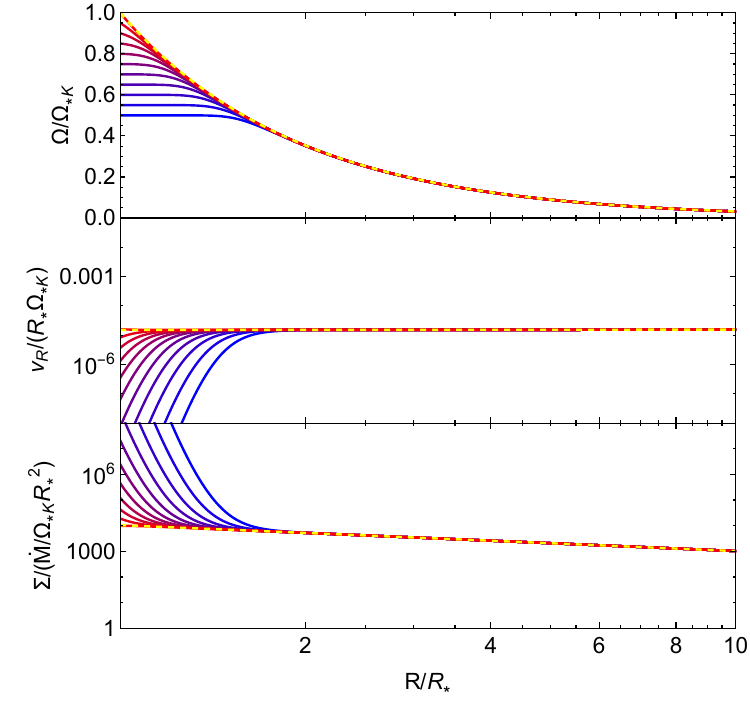}
     \caption{Steady state disc solutions with $H/R=0.1$ and $j=1$ (left, accretion disc), $j=10$ (middle, decretion disc) and $j=100$ (right, decretion disc).  Each panel shows the angular frequency, the radial velocity,  and the surface density from top to bottom.  The yellow dashed lines show the Keplerian disc solutions ($\Omega_*=\Omega_{\rm * K}$) given by equations~(\ref{omkep}), (\ref{vrkep}), and~(\ref{mdot}). The solid blue-red lines show solutions with a boundary layer and stellar rotation $\Omega_*=0.5$ (blue), 0.55, 0.6, 0.65 $0.7$, 0.75, $0.8$, 0.85, 0.9, 0.95  and $1.0 \,\Omega_{\rm K*}$ (red).  }
     \label{fig:example}
 \end{figure*}

\begin{figure}
    \centering
        \includegraphics[width=0.8\columnwidth]{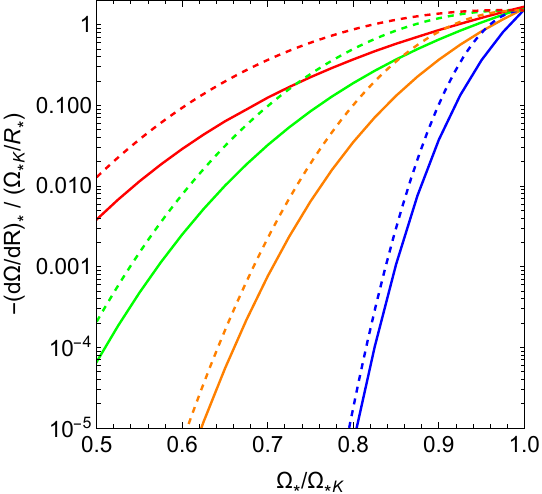}
    \caption{$d\Omega/ dR$ at $R=R_*$ for the decretion disc solutions with $j=10$. The disc aspect ratio is $H/R=0.05$ (blue), $0.1$ (yellow), $0.15$ (green) and 0.2 (red).  The solid lines show numerical solutions while the dashed lines show the analytic solutions given by equation~(\ref{domdr}).}
    \label{fig:omegacrit}
\end{figure}

\subsection{Numerical disc solutions}
\label{results}

The two coupled equations~(\ref{eqvr}) and~(\ref{eqom}) can be solved to  find the steady state $v_{\rm R}(R)$ and $\Omega(R)$. With the solution to these equations, the surface density can be found with equation~(\ref{mdot}). These equations are rather stiff and in order to find a solution using Mathematica, we include a "time" dependence into the equations so we solve for $\Omega=\Omega(R,\tau)$ and $v_{\rm  R}=v_{\rm R}(R,\tau)$. We change the derivatives with respect to $R$ to be partial and then we set the right hand sides of the equations to $\partial v_{\rm R}/\partial \tau$ and $-\partial \Omega/\partial \tau$, respectively, and then integrate until a steady state solution is reached in the variable $\tau$. Note that the variable $\tau$ is not time, it is just a way for us to begin with an initial guess and relax to the steady state solution.
In all of our models we take $R_{\rm out}=10\, R_*$. Since the boundary layer is radially narrow, the outer radius of the disc does not affect the solutions. Except where otherwise stated, we take the viscosity parameter to be $\alpha=0.1$. We check the convergence of our solutions by doubling the maximum step fraction and find that the relative difference between the solutions is $\lesssim 10^{-6}$.

Fig.~\ref{fig:example} shows the angular frequency, radial velocity, and surface density for $H/R=0.1$ with different values for the stellar spin $\Omega_*$. The left panel shows accretion disc solutions with $j=1$ and the middle and right panels show decretion disc solutions with $j=10$ and $j=100$, respectively. In the boundary layer region, where the flow differs from Keplerian, the angular frequency is lower than Keplerian frequency and smoothly transitions to the angular frequency of the stellar rotation at the inner boundary. The radial velocity in this region is smaller than the Keplerian radial velocity and this leads to a build up of material, as seen in the surface density. 
Note that the decretion solutions for $\Omega$ show little difference for $j=10$ and $j=100$.

While our solutions for $\Omega_* \gtrsim 0.7$ are in good agreement with \cite{Dong2021}, solutions for smaller values are somewhat different. The solutions in \cite{Dong2021} show a steep increase in $\Omega$ with radius in the boundary layer whereas our solutions have a flat profile. The timescale to reach a steady state solution in these cases may be prohibitively long. Fig.~\ref{fig:example} shows that the radial velocity in the boundary layer for small $\Omega_*$ is very small. Therefore in a time-dependent simulations, the steady state is not actually reached. The simulations in \cite{Dong2021} are run for a time of $t_{\rm end}=800\,P_{\rm *K}$ where $P_{\rm *K}=2\pi/\Omega_{* \rm K}$. Therefore in order to reach a steady state we need $R/|v_{\rm R}| \lesssim t_{\rm end}$ which requires $|v_{\rm R}| \gtrsim 10^{-4} R_* \Omega_{\rm *K}$. Fig.~\ref{fig:example} shows that the radial velocity for small $\Omega_*$ is too small for the steady state to be reached. 
Therefore the steady state solutions that we find here may only be realised for larger values of the stellar spin.

\subsection{What is $j$?}
\label{j}

Previous 1D simulations \citep[e.g.][]{Popham1991,Paczynski1991,Lee2013} found one unique value for $j$ for each spin rate $\Omega_*$. This is a result of their choice of  sound speed that depends upon the radial velocity. They constrained the disc aspect ratio at $R_*$ to find $j$.  However, because we chose $H/R$ to be constant, we have found that for any given spin rate, there is a disc solution with any value of $j$. Specifically, there are both accretion ($j=1$) and decretion solutions ($j \gg 1$) for the same spin rate.

Considering again the steady state Keplerian solutions in Section~\ref{Kep}, we see that for decretion disc solutions, the value of $j$ is related to $R_{\rm t}$ and this relates to the size of the disc and therefore the total mass of the disc.
In a time-dependent disc, material is injected close to the Be star equator. The material spreads outwards and builds up in the disc over time. 
The value of $\Sigma_*$ may be fixed by the stellar properties at $R_*$ (as we see in the next Section) but as the disc mass builds up, the  value for $j$ increases over time. 
A larger value for $j$ leads to a smaller radial velocity and a smaller $\dot M$.
This is in agreement with the 2D simulations in  \cite{Dong2021} who present two simulations for $\Omega_*=0.8$ with different initial outer disc radius, simulations Cs01A08 and Cs01A08L. The first has $R_{\rm out}=10\,R_*$ and they find $j=3.5$, while the second has $R_{\rm out}=21.2\,R_*$ and they find $j=66.9$. In the 2D time-dependent simulations, the outcome (accretion or decretion) is sensitive to  the choice of the initial disc surface density profile that provides an effective inflow. For a decretion disc to form, the decretion rate from the star must not be overwhelmed by the inflow rate. 
 
We note that the angular frequency profile $\Omega(R)$ for the decretion disc solutions is insensitive to changes in $j$ (see Fig.~\ref{fig:omegacrit}).   As we see in the next Section, the torque on the star depends upon $\Omega$ and the surface density at the inner disc edge, neither of which depend upon $j$. Therefore we can define a characteristic stellar spin rate which a star spins down to that is independent of $j$.

\subsection{Analytic approximations in the disc boundary layer}
\label{analytic}

We now make analytic approximations to the disc solutions. Fig.~\ref{fig:example} shows that in the boundary layer, if the steady state solution is reached, then for a range of stellar spins, there may be solutions that have $\Omega \approx \Omega_*$ in the boundary layer. With this approximation and in the limit that $c_{\rm s} \gg v_{\rm R}$, we can solve equation~(\ref{eqvr}) to find
\begin{equation}
    v_{\rm R}= v_{\rm R0} 
     \left(\frac{R}{R_*}\right)^\frac{1-h^2}{h^2}
    \exp{\left\{ -\frac{\Omega_*^2 }{3 h^2 \Omega_{\rm K}^2} \right\}},
    \label{vran}
\end{equation}
where $v_{\rm R0}$ is a constant of integration and $h=H/R$. We can approximate the constant by finding the Keplerian radial velocity with equation~(\ref{vrkep}) at the radius where $\Omega_{\rm K}=\Omega_*$ given by
\begin{equation}
    R_{\rm join}=\left(\frac{G M}{\Omega_*^2}\right)^{1/3}.
\end{equation}
The constant is
\begin{equation}
    v_{R0}= v_{\rm R,K}(R_{\rm join})
\left(\frac{R_{\rm join}}{R_*}\right)^{-\frac{1-h^2}{h^2}}
    \exp{\left\{ \frac{1 }{3 h^2  } \right\}},
\end{equation}
where the Keplerian radial velocity is given in equation~(\ref{vrkep}).
We can write equation~(\ref{vran}) as
\begin{equation}
    v_{\rm R}=  v_{\rm R,K}(R_{\rm join}) 
    \left(\frac{R}{R_{\rm join}}\right)^{\frac{1-h^2}{h^2}} \exp{\left\{ \frac{1}{3h^2}\left(1-\frac{\Omega_*^2 }{ \Omega_{\rm K}^2}\right) \right\}}.
    \label{vran2}
\end{equation}
With this approximation we can substitute into equation~(\ref{eqom}) to find
\begin{equation} 
    \left(\frac{d\Omega}{dR}\right)_* = -\frac{3C_{j}\Omega_{\rm *}}{2R_*}
\left(\frac{\Omega_*}{\Omega_{\rm *K}}\right)^{\frac{2-h^2}{3h^2}}
   \exp{\left\{ \frac{1}{3h^2}\left(1-\frac{\Omega_*^2 }{ \Omega_{\rm *K}^2}\right) \right\}},
   \label{domdr}
\end{equation}
where 
\begin{equation}
C_{j} =  \frac{ 1-j\left(\frac{ \Omega_{\rm * K}}{\Omega_*}\right)}{1-j\left(\frac{\Omega_{\rm * }}{\Omega_{\rm *K }}\right)^{1/3}}.
\end{equation}
In the limit $j\gg 1$, this is independent of $j$, as expected. The dashed lines in Fig.~\ref{fig:omegacrit} show the analytic solutions compared to the numerical solutions found in Section~\ref{results}. There is fairly good agreement between the analytic solution and the numerical solutions.  We use this analytic approximation in the next section where we calculate the torque on the star from the disc.

\section{Disc-star connection }
\label{discstar}

In this Section we  first calculate the torque on the star from the disc. Then we use rotating stellar models to find the surface density at the inner disc edge where it connects to the star. Finally, we find the characteristic stellar spin where the torque on the star becomes negligible, and the resulting decretion rates.

\begin{figure}
    \centering
    \includegraphics[width=0.9\linewidth]{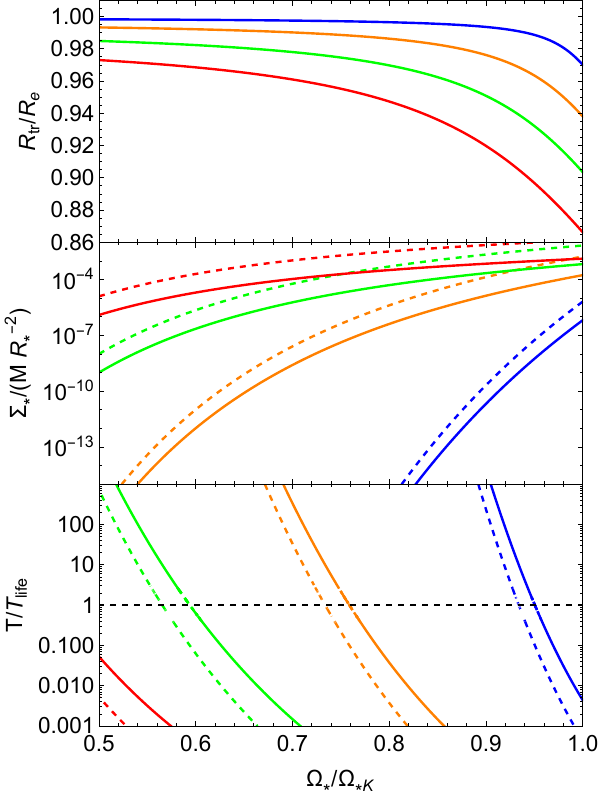}
    \caption{The connection between the star and the disc. Upper panel: The transition radius from the star to disc, $R_{\rm tr}$, in units of the equatorial radius versus the stellar spin rate. Middle panel: The surface density of the star at $R_{\rm tr}$. Lower panel: The timescale to change the stellar spin relative to the Be star lifetime.  In each panel, the disc aspect ratio is $H/R=0.05$ (blue), 0.1 (orange), 0.15 (green) and 0.2 (red). In the lower two panels, $X=0.01$ (solid lines) and $X=0.1$ (dashed lines).  }
    \label{fig:combine}
\end{figure}

\subsection{Torque on the star from the disc}

The spin of the star may evolve through two processes, first, the accretion of material with a different specific angular momentum to that of the star, and second, the action of the viscous torque in the disc on the star \citep{Frank2002}. Since (in general) the inner boundary condition should be that the star and the disc rotate at the same frequency at the stellar radius, accretion cannot change the spin of the star in our steady state models. The spin of the star can change through the torque from the disc \citep[e.g.][]{Glatzel1999}.

The viscous torque in the disc is 
\begin{equation}
G=2\pi R \nu\Sigma  R^2 \frac{d\Omega}{dR} 
\label{torque}
\end{equation}
\citep{Pringle1981}.
In the Keplerian steady state accretion disc solution (equation~\ref{Kepacc})  there is no torque on the star because $\Sigma=0$ at $R=R_*$. However, as shown in Fig.~\ref{fig:example}, this is not the case when boundary layer effects are taken into account or for decretion discs.  Fig.~\ref{fig:omegacrit} shows  $d\Omega/dR$ at the inner disc radius as a function of stellar rotation rate for different values of $H/R$.  This gradient is insensitive to $j$ for $j\gg1$. Note that for steady state accretion discs, $(d\Omega/dR)_*$ is always positive and the star always spins up. For steady state decretion discs, $(d\Omega/dR)_*$ is always negative the star always spins down. 

The Be star may get spun up through the accretion disc solutions during the evolution of the progenitor star of the neutron star companion. The Be star could even be spun up to breakup during this phase, but note that there is zero torque on the star from the disc at breakup. Once the accretion phase is over, the Be star can form a decretion disc and  the torque on the star now spins the star down. In order to calculate the torque on the star we need to evaluate equation~(\ref{torque}) at $R_*$ and this requires finding the surface density at the radius where the star and the disc connect.

\begin{figure}
    \centering
    \includegraphics[width=0.9\linewidth]{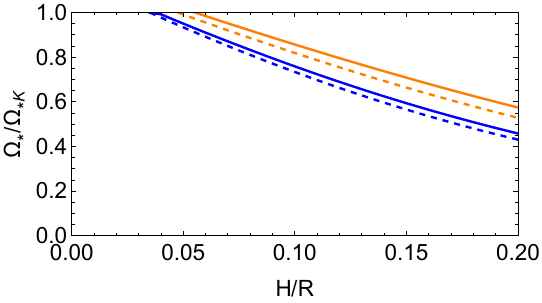}
    \caption{The characteristic stellar spin rate where the timescale for the torque to change the stellar spin becomes longer than the disc lifetime, $T=\,T_{\rm life}$. The solid lines are for $X=0.01$ and the dashed lines are for $X=0.1$. The blue lines are $\alpha=0.1$ and the orange lines are $\alpha=0.001$.}
    \label{fig:crit}
\end{figure}

\begin{figure}
    \centering
\includegraphics[width=0.9\columnwidth]{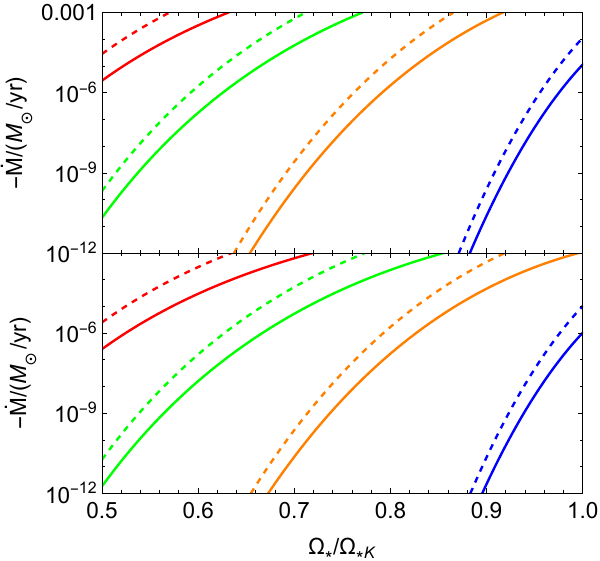}
    \caption{The decretion rate through the disc for $j=10$ (upper panel) and $j=100$ (lower panel). In each panel, the disc aspect ratio is $H/R=0.05$ (blue), 0.1 (orange), 0.15 (green) and 0.2 (red) while $X=0.01$ (solid lines) and $X=0.1$ (dashed lines). Note that these plots have $\alpha=0.1$ and the decretion rate scales with $\alpha$. }
    \label{fig:decrate}
\end{figure}

\subsection{Stellar model}

We constrain the surface density at the inner edge of the disc with a simple model of a rotating star. 
For an isolated star (without a disc), the star is in hydrostatic balance so that
\begin{equation}
    \frac{1}{\rho} \frac{\partial p}{\partial R}+\frac{G  M_* R}{(R^2+z^2)^{3/2}} = R \Omega_*^2,
    \label{hydro}
\end{equation}
where $p$ is the pressure and $\rho$ is the density.
In the outer layers of the star, we assume the star is uniformly rotating so that $\Omega_*=\,$const and the layers are isothermal and $p=c_{\rm s}^2 \rho$. 
The faster the star rotates, the more flattened its shape becomes. The height of the star above the equator follows
\begin{equation}
    z_0(R)=\left[ \left( \frac{1}{R_{\rm p}} -\frac{\Omega_*^2 R^2}{2 GM_*}\right)^{-2}-R^2 \right]^{1/2}
\end{equation}
\citep[e.g.][]{Paczynski1991}, where $R_{\rm p}=z_0(0)$ is the polar radius of the star that is found with
\begin{equation}
    R_{\rm p}=R_{\rm e}\left( 1+\frac{\Omega_*^2 R_{\rm e}^3}{2GM_*}\right)^{-1}
\end{equation}
and $R_{\rm e}$ is the radius at the equator. For a critically rotating star, $\Omega_*=(GM_*/R_{\rm e}^3)^{1/2}$ and  $R_{\rm p}=2/3\, R_{\rm e}$.

Now consider a star that has a disc. We define the transition radius where the star connects to the disc, $R_{\rm tr}$, to be where $z_0=H_*$, where $H_*$ is the disc height at $R=R_*$. In the limit $H/R\rightarrow 0$, the transition radius  $R_{\rm tr}\rightarrow R_{\rm e}$  and in this limit, the surface density at the inner disc edge $\Sigma_*=0$. The upper panel of Fig.~\ref{fig:combine} shows the transition radius for different stellar spins and disc aspect ratios.  The transition radius is smaller for larger disc aspect ratio and faster stellar spin rate. {\it We associate this transition radius with the radius of the star, $R_*$.}

We solve equation~(\ref{hydro}) in the limit $R\gg z$ to find
\begin{equation}
    \rho=\rho_{\rm in} \exp \left[\frac{G  M_* }{c_{\rm s}^2} \left(\frac{1}{R}-\frac{1}{R_{\rm in}}\right) +\frac{(R^2-R_{\rm in}^2) \Omega_*^2}{2c_{\rm s}^2} \right]
\end{equation}
with $c_{\rm s}=H_* \Omega_*$. The inner radius is taken to be $R_{\rm in}=0.9\,R_*$. We parameterize the constant with 
\begin{equation}
    \rho_{\rm in}= X \frac{M_*}{\frac{4}{3} \pi R_*^3} ,
\end{equation}
where $X$ is a dimensionless number that describes how the density at $R_{\rm in}$ compares to the average stellar density. For our Sun, $X= 0.019$ \citep{Bahcall2004}. We expect that the value is not too sensitive to the stellar mass. 
The core density at least is insensitive to stellar rotation \citep{Deupree2011}. Due to the uncertainty of the value for a rotating Be star, we consider two values, $X=0.01$ and $X=0.1$ and show that the results are not very sensitive to changes on this scale. 
The surface density in the outer stellar layers of the star is approximated with
\begin{equation}
    \Sigma = \int_{-z_0}^{z_0} \rho \, dz \approx 2 \rho z_0.
\end{equation}
The middle panel in  Fig.~\ref{fig:combine} shows the surface density at the transition radius that is larger for thicker discs and more rapidly rotating stars, as would be expected.

\subsection{Characteristic stellar spin rate}

The torque on the star is given in equation~(\ref{torque}) evaluated at the stellar surface $R=R_*$. The angular momentum of the star is approximated with
\begin{equation}
J_* = k M_* R_*^2 \Omega_*,
\end{equation}
where we take $k=0.076$ \citep[e.g.][]{Motz1952}
The timescale to change the spin of the star is
\begin{equation}
    T=\frac{J_*}{\frac{dJ_*}{dt}}=\frac{J_*}{G_*}.
    \label{timescale}
\end{equation}
The stellar spin evolves to the spin rate where the timsecale to change the spin is longer than the stellar lifetime.
The third panel in Fig.~\ref{fig:combine} shows the timescale to change the stellar spin. It is scaled by the lifetime of the star that we take to be $T_{\rm life}=3\times 10^6\,\rm yr$ for a star of mass $M_*=18\,\rm M_\odot$ with radius $R_*=8\,R_\odot$. While this is plotted for $j=10$, this is very insensitive to $j$. The stellar spin may decrease and evolve to be around the line $T/T_{\rm life}=1$ that is shown by the dashed black line.  Note that the steepness of the lines suggests that the Be stars can spin down rapidly initially and spend most of their lifetime in close to a constant rotation state.

Fig.~\ref{fig:crit} shows the spin rate where $T=T_{\rm life}$ for varying $H/R$. This critical spin rate is in good agreement with the 2D models of \cite{Dong2021} for $H/R=0.1$ and $H/R=0.15$. We also show the critical spin rate with a significantly smaller value for $\alpha$ and see that it isn't too sensitive to this parameter. We note that the trend of the critical spin rate with the disc aspect ratio (or temperature) may be in agreement with observations that suggest that lower mass (cooler) stars must rotate more rapidly to form a disc \citep{Huang2010}.

\subsection{Decretion rate}

Now that we have an estimate of the surface density at the inner edge of the disc, we can calculate the decretion rate with equation~(\ref{mdot}). Fig.~\ref{fig:decrate} shows the mass transfer rate through the disc for two different values of $j$. The decretion rates around $T \approx T_{\rm life}$ are consistent with observed rates in Be stars \citep[e.g.][]{Snow1981,Vieira2017}.  We note that the decretion rate scales with the viscosity parameter $\alpha$.
In the absence of any infalling material, decretion disc solutions are the only possible disc solutions.  However, if there is infall accretion on to the disc away from the inner boundary, this may drive accretion if it overwhelms the torque from the star.

\section{Conclusions and Discussion}
\label{concs}

Be stars are often found in binaries with an evolved companion. During the stellar evolution of the companion neutron star, there is an accretion phase on to the Be star during which the star may more than double in mass. This accretion phase leads to the spin up of the star. After the infall accretion subsides, we have shown that boundary layers effects in a geometrically thick accretion disc, that connects to a rotationally flattened star, can lead to the formation of a decretion disc around a Be star for rotation rates less than the stellar break up rate. The decretion rate is larger for faster stellar rotation and geometrically thicker discs. The decretion disc exerts a torque on the star that slows the spin. The torque is a strongly increasing function of the stellar spin. Therefore in practice the stellar spin approaches a critical value where the torque becomes negligible but decretion continues. For $H/R\approx 0.1$, the critical spin rate is around $0.7\, \Omega_{\rm *K}$. The critical spin rate decreases with increasing disc aspect ratio $H/R$.

We have assumed that the \cite{SS1973} $\alpha$ parameter is constant throughout the disc. However, the disc is only unstable to the magnetorotational instability where $d\Omega/dR<0$ \citep{BH1991}. For our decretion disc solutions, this condition is satisfied throughout the disc. In the accretion disc solutions, especially for the larger spin rates, the inner parts of the disc may not be MRI unstable. In this case,  angular momentum transport may be driven by alternate mechanisms such as acoustic waves generated by global supersonic shear instabilities and vortices \citep{Belyaev2012,Belyaev2013,Dittmann2021,Coleman2022,Coleman2022b,Fu2023,Fu2024,Dittmann2024}.

\section*{Acknowledgements}

We thank the anonymous referee for useful comments. We acknowledge support from NASA through grants 80NSSC21K0395, 80NSSC19K0443 and 80NSSC23M0104.

%%%%%%%%%%%%%%%%%%%%%%%%%%%%%%%%%%%%%%%%%%%%%%%%%%
\section*{Data Availability}

The data underlying this article will be shared on reasonable request to the corresponding author.



\bibliographystyle{mnras}
%\bibliography{mnras} % if your bibtex file is called example.bib

\begin{thebibliography}{}
\makeatletter
\relax
\def\mn@urlcharsother{\let\do\@makeother \do\$\do\&\do\#\do\^\do\_\do\%\do\~}
\def\mn@doi{\begingroup\mn@urlcharsother \@ifnextchar [ {\mn@doi@} {\mn@doi@[]}}
\def\mn@doi@[#1]#2{\def\@tempa{#1}\ifx\@tempa\@empty \href {http://dx.doi.org/#2} {doi:#2}\else \href {http://dx.doi.org/#2} {#1}\fi \endgroup}
\def\mn@eprint#1#2{\mn@eprint@#1:#2::\@nil}
\def\mn@eprint@arXiv#1{\href {http://arxiv.org/abs/#1} {{\tt arXiv:#1}}}
\def\mn@eprint@dblp#1{\href {http://dblp.uni-trier.de/rec/bibtex/#1.xml} {dblp:#1}}
\def\mn@eprint@#1:#2:#3:#4\@nil{\def\@tempa {#1}\def\@tempb {#2}\def\@tempc {#3}\ifx \@tempc \@empty \let \@tempc \@tempb \let \@tempb \@tempa \fi \ifx \@tempb \@empty \def\@tempb {arXiv}\fi \@ifundefined {mn@eprint@\@tempb}{\@tempb:\@tempc}{\expandafter \expandafter \csname mn@eprint@\@tempb\endcsname \expandafter{\@tempc}}}

\bibitem[\protect\citeauthoryear{{Artymowicz} \& {Lubow}}{{Artymowicz} \& {Lubow}}{1994}]{Artymowicz1994}
{Artymowicz} P.,  {Lubow} S.~H.,  1994, \mn@doi [ApJ] {10.1086/173679}, \href {http://adsabs.harvard.edu/abs/1994ApJ...421..651A} {421, 651}

\bibitem[\protect\citeauthoryear{{Bahcall} \& {Pinsonneault}}{{Bahcall} \& {Pinsonneault}}{2004}]{Bahcall2004}
{Bahcall} J.~N.,  {Pinsonneault} M.~H.,  2004, \mn@doi [\prl] {10.1103/PhysRevLett.92.121301}, \href {https://ui.adsabs.harvard.edu/abs/2004PhRvL..92l1301B} {92, 121301}

\bibitem[\protect\citeauthoryear{{Balbus} \& {Hawley}}{{Balbus} \& {Hawley}}{1991}]{BH1991}
{Balbus} S.~A.,  {Hawley} J.~F.,  1991, \mn@doi [ApJ] {10.1086/170270}, \href {http://adsabs.harvard.edu/abs/1991ApJ...376..214B} {376, 214}

\bibitem[\protect\citeauthoryear{{Belyaev} \& {Rafikov}}{{Belyaev} \& {Rafikov}}{2012}]{Belyaev2012}
{Belyaev} M.~A.,  {Rafikov} R.~R.,  2012, \mn@doi [\apj] {10.1088/0004-637X/752/2/115}, \href {https://ui.adsabs.harvard.edu/abs/2012ApJ...752..115B} {752, 115}

\bibitem[\protect\citeauthoryear{{Belyaev}, {Rafikov}  \& {Stone}}{{Belyaev} et~al.}{2013}]{Belyaev2013}
{Belyaev} M.~A.,  {Rafikov} R.~R.,   {Stone} J.~M.,  2013, \mn@doi [\apj] {10.1088/0004-637X/770/1/67}, \href {https://ui.adsabs.harvard.edu/abs/2013ApJ...770...67B} {770, 67}

\bibitem[\protect\citeauthoryear{{Bisnovatyi-Kogan}}{{Bisnovatyi-Kogan}}{1993}]{Bisnovatyi1993}
{Bisnovatyi-Kogan} G.~S.,  1993, \aap, \href {https://ui.adsabs.harvard.edu/abs/1993A&A...274..796B} {274, 796}

\bibitem[\protect\citeauthoryear{{Brandt} \& {Podsiadlowski}}{{Brandt} \& {Podsiadlowski}}{1995}]{Brandt1995}
{Brandt} N.,  {Podsiadlowski} P.,  1995, \mn@doi [\mnras] {10.1093/mnras/274.2.461}, \href {http://adsabs.harvard.edu/abs/1995MNRAS.274..461B} {274, 461}

\bibitem[\protect\citeauthoryear{{Carciofi} \& {Bjorkman}}{{Carciofi} \& {Bjorkman}}{2008}]{Carciofi2008}
{Carciofi} A.~C.,  {Bjorkman} J.~E.,  2008, \mn@doi [\apj] {10.1086/589875}, \href {https://ui.adsabs.harvard.edu/abs/2008ApJ...684.1374C} {684, 1374}

\bibitem[\protect\citeauthoryear{{Coleman}, {Rafikov}  \& {Philippov}}{{Coleman} et~al.}{2022a}]{Coleman2022b}
{Coleman} M. S.~B.,  {Rafikov} R.~R.,   {Philippov} A.~A.,  2022a, \mn@doi [\mnras] {10.1093/mnras/stab2962}, \href {https://ui.adsabs.harvard.edu/abs/2022MNRAS.509..440C} {509, 440}

\bibitem[\protect\citeauthoryear{{Coleman}, {Rafikov}  \& {Philippov}}{{Coleman} et~al.}{2022b}]{Coleman2022}
{Coleman} M. S.~B.,  {Rafikov} R.~R.,   {Philippov} A.~A.,  2022b, \mn@doi [\mnras] {10.1093/mnras/stac732}, \href {https://ui.adsabs.harvard.edu/abs/2022MNRAS.512.2945C} {512, 2945}

\bibitem[\protect\citeauthoryear{{Collins}}{{Collins}}{1987}]{Collins1987}
{Collins} II G.~W.,  1987, in {Slettebak} A.,  {Snow} T.~P.,  eds, IAU Colloq. 92: Physics of Be Stars. p.~3

\bibitem[\protect\citeauthoryear{{Colpi}, {Nannurelli}  \& {Calvani}}{{Colpi} et~al.}{1991}]{Colpi1991}
{Colpi} M.,  {Nannurelli} M.,   {Calvani} M.,  1991, \mn@doi [\mnras] {10.1093/mnras/253.1.55}, \href {https://ui.adsabs.harvard.edu/abs/1991MNRAS.253...55C} {253, 55}

\bibitem[\protect\citeauthoryear{{Cyr}, {Jones}, {Panoglou}, {Carciofi}  \& {Okazaki}}{{Cyr} et~al.}{2017}]{Cyr2017}
{Cyr} I.~H.,  {Jones} C.~E.,  {Panoglou} D.,  {Carciofi} A.~C.,   {Okazaki} A.~T.,  2017, \mn@doi [\mnras] {10.1093/mnras/stx1427}, \href {http://adsabs.harvard.edu/abs/2017MNRAS.471..596C} {471, 596}

\bibitem[\protect\citeauthoryear{{Deupree}}{{Deupree}}{2011}]{Deupree2011}
{Deupree} R.~G.,  2011, \mn@doi [\apj] {10.1088/0004-637X/735/2/69}, \href {https://ui.adsabs.harvard.edu/abs/2011ApJ...735...69D} {735, 69}

\bibitem[\protect\citeauthoryear{{Dittmann}}{{Dittmann}}{2021}]{Dittmann2021}
{Dittmann} A.~J.,  2021, \mn@doi [\mnras] {10.1093/mnras/stab2682}, \href {https://ui.adsabs.harvard.edu/abs/2021MNRAS.508.1842D} {508, 1842}

\bibitem[\protect\citeauthoryear{{Dittmann}}{{Dittmann}}{2024}]{Dittmann2024}
{Dittmann} A.~J.,  2024, \mn@doi [arXiv e-prints] {10.48550/arXiv.2405.20367}, \href {https://ui.adsabs.harvard.edu/abs/2024arXiv240520367D} {p. arXiv:2405.20367}

\bibitem[\protect\citeauthoryear{{Dodd}, {Oudmaijer}, {Radley}, {Vioque}  \& {Frost}}{{Dodd} et~al.}{2024}]{Dodd2024}
{Dodd} J.~M.,  {Oudmaijer} R.~D.,  {Radley} I.~C.,  {Vioque} M.,   {Frost} A.~J.,  2024, \mn@doi [\mnras] {10.1093/mnras/stad3105}, \href {https://ui.adsabs.harvard.edu/abs/2024MNRAS.527.3076D} {527, 3076}

\bibitem[\protect\citeauthoryear{{Dong}, {Jiang}  \& {Armitage}}{{Dong} et~al.}{2021}]{Dong2021}
{Dong} J.,  {Jiang} Y.-F.,   {Armitage} P.~J.,  2021, \mn@doi [\apj] {10.3847/1538-4357/ac1941}, \href {https://ui.adsabs.harvard.edu/abs/2021ApJ...921...54D} {921, 54}

\bibitem[\protect\citeauthoryear{{Franchini} \& {Martin}}{{Franchini} \& {Martin}}{2019}]{Franchini2019bestars}
{Franchini} A.,  {Martin} R.~G.,  2019, \mn@doi [\apjl] {10.3847/2041-8213/ab3920}, \href {https://ui.adsabs.harvard.edu/abs/2019ApJ...881L..32F} {881, L32}

\bibitem[\protect\citeauthoryear{{Franchini} \& {Martin}}{{Franchini} \& {Martin}}{2021}]{Franchini2021}
{Franchini} A.,  {Martin} R.~G.,  2021, \mn@doi [\apjl] {10.3847/2041-8213/ac4029}, \href {https://ui.adsabs.harvard.edu/abs/2021ApJ...923L..18F} {923, L18}

\bibitem[\protect\citeauthoryear{{Frank}, {King}  \& {Raine}}{{Frank} et~al.}{2002}]{Frank2002}
{Frank} J.,  {King} A.,   {Raine} D.~J.,  2002, {Accretion Power in Astrophysics: Third Edition}

\bibitem[\protect\citeauthoryear{{Fu}, {Huang}  \& {Yu}}{{Fu} et~al.}{2023}]{Fu2023}
{Fu} Z.,  {Huang} S.,   {Yu} C.,  2023, \mn@doi [\apj] {10.3847/1538-4357/acac9c}, \href {https://ui.adsabs.harvard.edu/abs/2023ApJ...945..165F} {945, 165}

\bibitem[\protect\citeauthoryear{{Fu}, {Huang}  \& {Yu}}{{Fu} et~al.}{2024}]{Fu2024}
{Fu} Z.,  {Huang} S.,   {Yu} C.,  2024, \mn@doi [\apj] {10.3847/1538-4357/ad7584}, \href {https://ui.adsabs.harvard.edu/abs/2024ApJ...975...80F} {975, 80}

\bibitem[\protect\citeauthoryear{{Glatzel} \& {Obach}}{{Glatzel} \& {Obach}}{1999}]{Glatzel1999}
{Glatzel} W.,  {Obach} C.,  1999, \mn@doi [\mnras] {10.1046/j.1365-8711.1999.02725.x}, \href {https://ui.adsabs.harvard.edu/abs/1999MNRAS.308..147G} {308, 147}

\bibitem[\protect\citeauthoryear{{Hanuschik}}{{Hanuschik}}{1996}]{Hanuschik1996}
{Hanuschik} R.~W.,  1996, A\&A, \href {http://adsabs.harvard.edu/abs/1996A%26A...308..170H} {308, 170}

\bibitem[\protect\citeauthoryear{{Huang}, {Gies}  \& {McSwain}}{{Huang} et~al.}{2010}]{Huang2010}
{Huang} W.,  {Gies} D.~R.,   {McSwain} M.~V.,  2010, \mn@doi [\apj] {10.1088/0004-637X/722/1/605}, \href {https://ui.adsabs.harvard.edu/abs/2010ApJ...722..605H} {722, 605}

\bibitem[\protect\citeauthoryear{{Hummel}}{{Hummel}}{1998}]{Hummel1998}
{Hummel} W.,  1998, \aap, \href {http://adsabs.harvard.edu/abs/1998A%26A...330..243H} {330, 243}

\bibitem[\protect\citeauthoryear{{Jones}, {Sigut}  \& {Porter}}{{Jones} et~al.}{2008}]{Jones2008}
{Jones} C.~E.,  {Sigut} T.~A.~A.,   {Porter} J.~M.,  2008, \mn@doi [\mnras] {10.1111/j.1365-2966.2008.13206.x}, \href {http://adsabs.harvard.edu/abs/2008MNRAS.386.1922J} {386, 1922}

\bibitem[\protect\citeauthoryear{{Lau}, {Potter}  \& {Tout}}{{Lau} et~al.}{2011}]{Lau2011}
{Lau} H. H.~B.,  {Potter} A.~T.,   {Tout} C.~A.,  2011, \mn@doi [\mnras] {10.1111/j.1365-2966.2011.18766.x}, \href {https://ui.adsabs.harvard.edu/abs/2011MNRAS.415..959L} {415, 959}

\bibitem[\protect\citeauthoryear{{Lee}}{{Lee}}{2013}]{Lee2013}
{Lee} U.,  2013, \mn@doi [\pasj] {10.1093/pasj/65.6.122}, \href {https://ui.adsabs.harvard.edu/abs/2013PASJ...65..122L} {65, 122}

\bibitem[\protect\citeauthoryear{{Lee}, {Osaki}  \& {Saio}}{{Lee} et~al.}{1991}]{Lee1991}
{Lee} U.,  {Osaki} Y.,   {Saio} H.,  1991, MNRAS, \href {http://adsabs.harvard.edu/abs/1991MNRAS.250..432L} {250, 432}

\bibitem[\protect\citeauthoryear{{Lubow}, {Martin}  \& {Nixon}}{{Lubow} et~al.}{2015}]{Lubowetal2015}
{Lubow} S.~H.,  {Martin} R.~G.,   {Nixon} C.,  2015, \mn@doi [ApJ] {10.1088/0004-637X/800/2/96}, \href {http://adsabs.harvard.edu/abs/2015ApJ...800...96L} {800, 96}

\bibitem[\protect\citeauthoryear{{Martin}, {Tout}  \& {Pringle}}{{Martin} et~al.}{2009}]{Martinetal2009b}
{Martin} R.~G.,  {Tout} C.~A.,   {Pringle} J.~E.,  2009, \mn@doi [MNRAS] {10.1111/j.1365-2966.2009.15031.x}, \href {http://adsabs.harvard.edu/abs/2009MNRAS.397.1563M} {397, 1563}

\bibitem[\protect\citeauthoryear{{Martin}, {Nixon}, {Armitage}, {Lubow}  \& {Price}}{{Martin} et~al.}{2014}]{Martinetal2014}
{Martin} R.~G.,  {Nixon} C.,  {Armitage} P.~J.,  {Lubow} S.~H.,   {Price} D.~J.,  2014, \mn@doi [ApJL] {10.1088/2041-8205/790/2/L34}, \href {http://adsabs.harvard.edu/abs/2014ApJ...790L..34M} {790, L34}

\bibitem[\protect\citeauthoryear{{Martin}, {Lubow}, {Armitage}  \& {Price}}{{Martin} et~al.}{2024}]{Martin2024}
{Martin} R.~G.,  {Lubow} S.~H.,  {Armitage} P.~J.,   {Price} D.~J.,  2024, \mn@doi [\mnras] {10.1093/mnras/stae1143}, \href {https://ui.adsabs.harvard.edu/abs/2024MNRAS.530.4148M} {530, 4148}

\bibitem[\protect\citeauthoryear{{McSwain} \& {Gies}}{{McSwain} \& {Gies}}{2005}]{McSwain2005}
{McSwain} M.~V.,  {Gies} D.~R.,  2005, \mn@doi [\apjs] {10.1086/432757}, \href {https://ui.adsabs.harvard.edu/abs/2005ApJS..161..118M} {161, 118}

\bibitem[\protect\citeauthoryear{{Mermilliod}}{{Mermilliod}}{1982}]{Mermilliod1982}
{Mermilliod} J.~C.,  1982, \aap, \href {https://ui.adsabs.harvard.edu/abs/1982A&A...109...48M} {109, 48}

\bibitem[\protect\citeauthoryear{{Miranda} \& {Lai}}{{Miranda} \& {Lai}}{2015}]{Miranda2015}
{Miranda} R.,  {Lai} D.,  2015, \mn@doi [\mnras] {10.1093/mnras/stv1450}, \href {http://adsabs.harvard.edu/abs/2015MNRAS.452.2396M} {452, 2396}

\bibitem[\protect\citeauthoryear{{Motz}}{{Motz}}{1952}]{Motz1952}
{Motz} L.,  1952, \mn@doi [\apj] {10.1086/145570}, \href {https://ui.adsabs.harvard.edu/abs/1952ApJ...115..562M} {115, 562}

\bibitem[\protect\citeauthoryear{Nathaniel, Langer, Simón-Díaz, Holgado, de Burgos  \& Hastings}{Nathaniel et~al.}{2025}]{Nathaniel2025}
Nathaniel K.,  Langer N.,  Simón-Díaz S.,  Holgado G.,  de Burgos A.,   Hastings B.,  2025, Spindown of massive main sequence stars in the Milky Way (\mn@eprint {arXiv} {2502.12107}), \url {https://arxiv.org/abs/2502.12107}

\bibitem[\protect\citeauthoryear{{Nixon} \& {Pringle}}{{Nixon} \& {Pringle}}{2020}]{Nixon2020}
{Nixon} C.~J.,  {Pringle} J.~E.,  2020, \mn@doi [\apjl] {10.3847/2041-8213/abd17e}, \href {https://ui.adsabs.harvard.edu/abs/2020ApJ...905L..29N} {905, L29}

\bibitem[\protect\citeauthoryear{{Nixon} \& {Pringle}}{{Nixon} \& {Pringle}}{2021}]{Nixon2021}
{Nixon} C.~J.,  {Pringle} J.~E.,  2021, \mn@doi [\na] {10.1016/j.newast.2020.101493}, \href {https://ui.adsabs.harvard.edu/abs/2021NewA...8501493N} {85, 101493}

\bibitem[\protect\citeauthoryear{{Okazaki}}{{Okazaki}}{2016}]{Okazaki2016}
{Okazaki} A.~T.,  2016, in {Sigut} T.~A.~A.,  {Jones} C.~E.,  eds,  Astronomical Society of the Pacific Conference Series Vol. 506, Bright Emissaries: Be Stars as Messengers of Star-Disk Physics. p.~3

\bibitem[\protect\citeauthoryear{{Okazaki} \& {Negueruela}}{{Okazaki} \& {Negueruela}}{2001}]{Okazaki2001}
{Okazaki} A.~T.,  {Negueruela} I.,  2001, \mn@doi [A\&A] {10.1051/0004-6361:20011083}, \href {http://adsabs.harvard.edu/abs/2001A%26A...377..161O} {377, 161}

\bibitem[\protect\citeauthoryear{{Okazaki}, {Bate}, {Ogilvie}  \& {Pringle}}{{Okazaki} et~al.}{2002}]{Okazaki2002}
{Okazaki} A.~T.,  {Bate} M.~R.,  {Ogilvie} G.~I.,   {Pringle} J.~E.,  2002, \mn@doi [MNRAS] {10.1046/j.1365-8711.2002.05960.x}, \href {http://adsabs.harvard.edu/abs/2002MNRAS.337..967O} {337, 967}

\bibitem[\protect\citeauthoryear{{Overton}, {Martin}, {Lubow}  \& {Lepp}}{{Overton} et~al.}{2024}]{Overton2024}
{Overton} M.,  {Martin} R.~G.,  {Lubow} S.~H.,   {Lepp} S.,  2024, \mn@doi [\mnras] {10.1093/mnrasl/slad172}, \href {https://ui.adsabs.harvard.edu/abs/2024MNRAS.528L.106O} {528, L106}

\bibitem[\protect\citeauthoryear{{Paczynski}}{{Paczynski}}{1991}]{Paczynski1991}
{Paczynski} B.,  1991, \mn@doi [\apj] {10.1086/169846}, \href {https://ui.adsabs.harvard.edu/abs/1991ApJ...370..597P} {370, 597}

\bibitem[\protect\citeauthoryear{{Panoglou}, {Faes}, {Carciofi}, {Okazaki}, {Baade}, {Rivinius}  \& {Borges Fernandes}}{{Panoglou} et~al.}{2018}]{Panoglou2018}
{Panoglou} D.,  {Faes} D.~M.,  {Carciofi} A.~C.,  {Okazaki} A.~T.,  {Baade} D.,  {Rivinius} T.,   {Borges Fernandes} M.,  2018, \mn@doi [\mnras] {10.1093/mnras/stx2497}, \href {http://adsabs.harvard.edu/abs/2018MNRAS.473.3039P} {473, 3039}

\bibitem[\protect\citeauthoryear{{Papaloizou} \& {Pringle}}{{Papaloizou} \& {Pringle}}{1977}]{Papaloizou1977}
{Papaloizou} J.,  {Pringle} J.~E.,  1977, \mn@doi [\mnras] {10.1093/mnras/181.3.441}, \href {https://ui.adsabs.harvard.edu/abs/1977MNRAS.181..441P} {181, 441}

\bibitem[\protect\citeauthoryear{{Pols}, {Cote}, {Waters}  \& {Heise}}{{Pols} et~al.}{1991}]{Pols1991}
{Pols} O.~R.,  {Cote} J.,  {Waters} L.~B.~F.~M.,   {Heise} J.,  1991, \aap, \href {https://ui.adsabs.harvard.edu/abs/1991A&A...241..419P} {241, 419}

\bibitem[\protect\citeauthoryear{{Popham} \& {Narayan}}{{Popham} \& {Narayan}}{1991}]{Popham1991}
{Popham} R.,  {Narayan} R.,  1991, \mn@doi [\apj] {10.1086/169847}, \href {https://ui.adsabs.harvard.edu/abs/1991ApJ...370..604P} {370, 604}

\bibitem[\protect\citeauthoryear{{Popham}, {Narayan}, {Hartmann}  \& {Kenyon}}{{Popham} et~al.}{1993}]{Popham1993}
{Popham} R.,  {Narayan} R.,  {Hartmann} L.,   {Kenyon} S.,  1993, \mn@doi [\apjl] {10.1086/187049}, \href {https://ui.adsabs.harvard.edu/abs/1993ApJ...415L.127P} {415, L127}

\bibitem[\protect\citeauthoryear{{Porter}}{{Porter}}{1996}]{Porter1996}
{Porter} J.~M.,  1996, MNRAS, \href {http://adsabs.harvard.edu/abs/1996MNRAS.280L..31P} {280, L31}

\bibitem[\protect\citeauthoryear{{Porter} \& {Rivinius}}{{Porter} \& {Rivinius}}{2003}]{Porter2003}
{Porter} J.~M.,  {Rivinius} T.,  2003, \mn@doi [\pasp] {10.1086/378307}, \href {http://adsabs.harvard.edu/abs/2003PASP..115.1153P} {115, 1153}

\bibitem[\protect\citeauthoryear{{Pringle}}{{Pringle}}{1977}]{Pringle1977}
{Pringle} J.~E.,  1977, \mn@doi [\mnras] {10.1093/mnras/178.2.195}, \href {https://ui.adsabs.harvard.edu/abs/1977MNRAS.178..195P} {178, 195}

\bibitem[\protect\citeauthoryear{{Pringle}}{{Pringle}}{1981}]{Pringle1981}
{Pringle} J.~E.,  1981, \mn@doi [ARA\&A] {10.1146/annurev.aa.19.090181.001033}, \href {http://adsabs.harvard.edu/abs/1981ARA%26A..19..137P} {19, 137}

\bibitem[\protect\citeauthoryear{{Pringle}}{{Pringle}}{1991}]{Pringle1991}
{Pringle} J.~E.,  1991, MNRAS, \href {http://adsabs.harvard.edu/abs/1991MNRAS.248..754P} {248, 754}

\bibitem[\protect\citeauthoryear{{Pringle} \& {Rees}}{{Pringle} \& {Rees}}{1972}]{Pringle1972}
{Pringle} J.~E.,  {Rees} M.~J.,  1972, \aap, \href {https://ui.adsabs.harvard.edu/abs/1972A&A....21....1P} {21, 1}

\bibitem[\protect\citeauthoryear{{Pringle} \& {Savonije}}{{Pringle} \& {Savonije}}{1979}]{Pringle1979}
{Pringle} J.~E.,  {Savonije} G.~J.,  1979, \mn@doi [\mnras] {10.1093/mnras/187.4.777}, \href {https://ui.adsabs.harvard.edu/abs/1979MNRAS.187..777P} {187, 777}

\bibitem[\protect\citeauthoryear{{Quirrenbach} et~al.,}{{Quirrenbach} et~al.}{1997}]{Quirrenbach1997}
{Quirrenbach} A.,  et~al., 1997, \mn@doi [\apj] {10.1086/303854}, \href {http://adsabs.harvard.edu/abs/1997ApJ...479..477Q} {479, 477}

\bibitem[\protect\citeauthoryear{{Raguzova} \& {Popov}}{{Raguzova} \& {Popov}}{2005}]{Raguzova2005}
{Raguzova} N.~V.,  {Popov} S.~B.,  2005, \mn@doi [Astronomical and Astrophysical Transactions] {10.1080/10556790500497311}, \href {https://ui.adsabs.harvard.edu/abs/2005A&AT...24..151R} {24, 151}

\bibitem[\protect\citeauthoryear{{Rast}, {Jones}, {Carciofi}, {Suffak}, {Fonseca Silva}, {Henry}  \& {Tycner}}{{Rast} et~al.}{2024}]{Rast2024}
{Rast} R.~G.,  {Jones} C.~E.,  {Carciofi} A.~C.,  {Suffak} M.~W.,  {Fonseca Silva} A.~C.,  {Henry} G.~W.,   {Tycner} C.,  2024, \mn@doi [\apj] {10.3847/1538-4357/ad40a2}, \href {https://ui.adsabs.harvard.edu/abs/2024ApJ...968...30R} {968, 30}

\bibitem[\protect\citeauthoryear{{Regev}}{{Regev}}{1983}]{Regev1983}
{Regev} O.,  1983, \aap, \href {https://ui.adsabs.harvard.edu/abs/1983A&A...126..146R} {126, 146}

\bibitem[\protect\citeauthoryear{{Regev} \& {Bertout}}{{Regev} \& {Bertout}}{1995}]{Regev1995}
{Regev} O.,  {Bertout} C.,  1995, \mn@doi [\mnras] {10.1093/mnras/272.1.71}, \href {https://ui.adsabs.harvard.edu/abs/1995MNRAS.272...71R} {272, 71}

\bibitem[\protect\citeauthoryear{{Reig}}{{Reig}}{2011}]{Reig2011}
{Reig} P.,  2011, \mn@doi [\apss] {10.1007/s10509-010-0575-8}, \href {http://adsabs.harvard.edu/abs/2011Ap%26SS.332....1R} {332, 1}

\bibitem[\protect\citeauthoryear{{Rivinius}, {Carciofi}  \& {Martayan}}{{Rivinius} et~al.}{2013}]{Rivinius2013}
{Rivinius} T.,  {Carciofi} A.~C.,   {Martayan} C.,  2013, \mn@doi [\aapr] {10.1007/s00159-013-0069-0}, \href {http://adsabs.harvard.edu/abs/2013A%26ARv..21...69R} {21, 69}

\bibitem[\protect\citeauthoryear{{Salvesen} \& {Pokawanvit}}{{Salvesen} \& {Pokawanvit}}{2020}]{Salvesen2020}
{Salvesen} G.,  {Pokawanvit} S.,  2020, \mn@doi [\mnras] {10.1093/mnras/staa1094}, \href {https://ui.adsabs.harvard.edu/abs/2020MNRAS.495.2179S} {495, 2179}

\bibitem[\protect\citeauthoryear{{Shakura} \& {Sunyaev}}{{Shakura} \& {Sunyaev}}{1973}]{SS1973}
{Shakura} N.~I.,  {Sunyaev} R.~A.,  1973, A\&A, \href {http://adsabs.harvard.edu/abs/1973A%26A....24..337S} {24, 337}

\bibitem[\protect\citeauthoryear{{Slettebak}}{{Slettebak}}{1982}]{Slettebak1982}
{Slettebak} A.,  1982, \mn@doi [ApJs] {10.1086/190820}, \href {http://adsabs.harvard.edu/abs/1982ApJS...50...55S} {50, 55}

\bibitem[\protect\citeauthoryear{{Slettebak}}{{Slettebak}}{1985}]{Slettebak1985}
{Slettebak} A.,  1985, \mn@doi [\apjs] {10.1086/191084}, \href {https://ui.adsabs.harvard.edu/abs/1985ApJS...59..769S} {59, 769}

\bibitem[\protect\citeauthoryear{{Snow}}{{Snow}}{1981}]{Snow1981}
{Snow} T.~P. J.,  1981, \mn@doi [\apj] {10.1086/159448}, \href {https://ui.adsabs.harvard.edu/abs/1981ApJ...251..139S} {251, 139}

\bibitem[\protect\citeauthoryear{{Suffak}, {Jones}  \& {Carciofi}}{{Suffak} et~al.}{2022}]{Suffak2022}
{Suffak} M.,  {Jones} C.~E.,   {Carciofi} A.~C.,  2022, \mn@doi [\mnras] {10.1093/mnras/stab3024}, \href {https://ui.adsabs.harvard.edu/abs/2022MNRAS.509..931S} {509, 931}

\bibitem[\protect\citeauthoryear{{Vieira}, {Carciofi}, {Bjorkman}, {Rivinius}, {Baade}  \& {R{\'\i}mulo}}{{Vieira} et~al.}{2017}]{Vieira2017}
{Vieira} R.~G.,  {Carciofi} A.~C.,  {Bjorkman} J.~E.,  {Rivinius} T.,  {Baade} D.,   {R{\'\i}mulo} L.~R.,  2017, \mn@doi [\mnras] {10.1093/mnras/stw2542}, \href {https://ui.adsabs.harvard.edu/abs/2017MNRAS.464.3071V} {464, 3071}

\bibitem[\protect\citeauthoryear{{Wheelwright}, {Bjorkman}, {Oudmaijer}, {Carciofi}, {Bjorkman}  \& {Porter}}{{Wheelwright} et~al.}{2012}]{Wheelwright2012}
{Wheelwright} H.~E.,  {Bjorkman} J.~E.,  {Oudmaijer} R.~D.,  {Carciofi} A.~C.,  {Bjorkman} K.~S.,   {Porter} J.~M.,  2012, \mn@doi [\mnras] {10.1111/j.1745-3933.2012.01241.x}, \href {https://ui.adsabs.harvard.edu/abs/2012MNRAS.423L..11W} {423, L11}

\bibitem[\protect\citeauthoryear{{de Mink}, {Langer}, {Izzard}, {Sana}  \& {de Koter}}{{de Mink} et~al.}{2013}]{deMink2013}
{de Mink} S.~E.,  {Langer} N.,  {Izzard} R.~G.,  {Sana} H.,   {de Koter} A.,  2013, \mn@doi [\apj] {10.1088/0004-637X/764/2/166}, \href {https://ui.adsabs.harvard.edu/abs/2013ApJ...764..166D} {764, 166}

\bibitem[\protect\citeauthoryear{{ud-Doula}, {Owocki}  \& {Kee}}{{ud-Doula} et~al.}{2018}]{udDoula2018}
{ud-Doula} A.,  {Owocki} S.~P.,   {Kee} N.~D.,  2018, \mn@doi [\mnras] {10.1093/mnras/sty1228}, \href {https://ui.adsabs.harvard.edu/abs/2018MNRAS.478.3049U} {478, 3049}

\makeatother
\end{thebibliography}

%\begin{thebibliography}{99}
%\end{thebibliography}

%%%%%%%%%%%%%%%%%%%%%%%%%%%%%%%%%%%%%%%%%%%%%%%%%%

%%%%%%%%%%%%%%%%% APPENDICES %%%%%%%%%%%%%%%%%%%%%

%%%%%%%%%%%%%%%%%%%%%%%%%%%%%%%%%%%%%%%%%%%%%%%%%%

% End of mnras_template.tex

% Don't change these lines
\bsp	% typesetting comment
\label{lastpage}
\end{document}